\documentclass[pre,aps,twocolumn,showpacs]{revtex4}
\usepackage{amsmath,amssymb,graphicx}
\usepackage{epsfig}
\usepackage{textcomp}
\usepackage{color}

\begin{document}
\title{Velocity fluctuations of population fronts propagating into metastable states}
\author{Baruch Meerson$^1$, Pavel V. Sasorov$^2$, and Yitzhak Kaplan$^1$}
\affiliation{$^1$Racah Institute of Physics, Hebrew University of
Jerusalem, Jerusalem 91904, Israel}
\affiliation{$^2$Institute for Theoretical and Experimental
Physics, Moscow 117218, Russia}

\pacs{05.40.-a, 05.10.Gg, 87.23.Cc, 02.50.Ga}
\begin{abstract}

The position of propagating population fronts fluctuates because of the discreteness of the individuals
and stochastic character of processes of birth, death and migration.  Here
we consider a Markov model of a population front propagating into a metastable state, and  focus on the weak noise limit. For typical, small fluctuations
the front motion is diffusive, and we calculate the front diffusion coefficient.
We also determine the probability distribution of rare, large fluctuations of the front position and, for a given average front velocity, find the most likely population density profile of the front. Implications of the theory for population extinction risk are briefly considered.
\end{abstract}
\maketitle

\section{Introduction}
Population fronts are ubiquitous in nature. They
control the speed of a multitude of processes ranging from epidemic outbursts to invasion of species to biological
evolution \cite{Murray,Petrovskii}. There is a fundamental difference
between fronts propagating into an unstable state and those propagating into a metastable state (a state which is linearly stable but  non-linearly unstable).
The former can be extremely sensitive to intrinsic noise in the leading edge
\cite{Tsimring,Derrida1,vanSaarloos03,Panja,Derrida06}. In particular, for pulled fronts, the front diffusion coefficient $D_f$, describing wandering of the front position around the position, predicted by the deterministic rate equation, is only \emph{logarithmically} small with $N$, the number of particles in the front region \cite{Derrida06}. Fronts propagating into metastable states are expected to behave ``normally" in this respect, and $D_f \sim 1/N$ scaling  was conjectured \cite{vanSaarloos03,Panja}. Surprisingly, not much is known beyond this conjecture. Yet fronts, propagating into metastable states, appear in many applications \cite{Murray,Petrovskii,Mikhailov}. For example, they describe colonization and extinction
waves when an Allee effect is present in the form of a critical population size needed for establishment of a local population \cite{Allee,Petrovskii}.  Motivated by this and other front propagation phenomena -- in physics, chemistry and biology -- we present here a systematic theory of fluctuations of the position of a population front propagating into a metastable state.

\section{Model}

Let a single population of individuals (we will call them particles) reside on a one-dimensional lattice of sites labeled by index $i$. The population size $n_i$ at each site varies in time as a result of two types of Markov processes \cite{Gardiner}. The first type involves  stochastic births and deaths with
rates $\lambda(n_i)$ and $\mu(n_i)$, respectively.
The second one is random walk of each particle between neighboring sites with rate coefficient $D_0$.

We assume that the typical population size scales as a large parameter, $K\gg 1$. Then for $n_i\gg 1$  one can write, in the leading order \cite{Doering,AM2010}:
\begin{equation}\label{rates}
    \lambda(n_i)=\nu K \bar{\lambda}(q_i)\;\;\;\;\;\mbox{and} \;\;\;\;\;\mu(n_i)=\nu K \bar{\mu}(q_i),
\end{equation}
where $q_i=n_i/K$ is the rescaled population size at site $i$, $\bar{\lambda}(q_i)\sim \bar{\mu}(q_i) \sim 1$, and $\nu$ is a rate coefficient. Assuming also a fast diffusion, $D_0\gg \nu$, we can use a continuous position coordinate $x$ instead of the discrete index $i$.
Neglecting for a moment the intrinsic noise, we arrive at a deterministic reaction-diffusion equation
\begin{equation}\label{rateeq3}
   \partial_t q=f(q)+\partial_x^2 q\,,
\end{equation}
where $f(q)=\bar{\lambda}(q) - \bar{\mu}(q)$, $D=D_0 h^2$ is the diffusion coefficient of the random walk, and $h$ is the lattice constant. The position coordinate $x$ in Eq.~(\ref{rateeq3}) is rescaled by the characteristic diffusion length $l_D=(D/\nu)^{1/2}$, and time is rescaled by $1/\nu$. We will restore the dimensions in our main results by recalling that the front velocity is measured in the units of $\nu l_D=(\nu D)^{1/2}$, and
the front diffusion coefficient $D_f$ is measured in the units of $D$.

A (deterministic) population front corresponds to a traveling wave solution (TWS) of Eq.~(\ref{rateeq3}), $q(x,t)=q(\xi)$, $\xi=x-c_0 t$, which interpolates between two homogeneous states described by zeros of
$f(q)$. Let one of them,  $q=q_*>0$, be located at $x\to -\infty$, whereas
the other, $q=0$, at $x\to \infty$. Each of them is linearly stable: $f^{\prime}(q_*)<0$ and $f^{\prime}(0)<0$. There is also a linearly unstable state $q_u$, so that $0<q_u<q_*$.  The deterministic TWS $q_0(\xi)$ obeys the equation
\begin{equation}\label{4}
q_0^{\prime\prime}+c_0 q_0^{\prime}+f(q_0)=0\,.
\end{equation}
This TWS is unique up to an arbitrary shift in $\xi$. The formal definition of the metastable state in this work is based on the effective potential $V(q)=\int_0^q f(q^{\prime}) \,dq^{\prime}$ of the deterministic problem \cite{Murray,Mikhailov,Petrovskii}.
The state $q=q_*>0$ is advancing ($c_0>0$) when $V(q_*)>0$.  In this case $q=0$ is called the metastable state. For $V(q_*)<0$ the state $q=q_*$ is
retreating ($c_0<0$), and so it is metastable. Finally, for $V(q_*)=0$ the front is standing, $c_0=0$, and so the two states $q=q_*$ and $q=0$ coexist.  The typical front width is on the order of $l_D$.

Now we go back to the discrete-lattice model and describe \textit{stochastic} dynamics of the population. Introduce the multivariate probability distribution
$P(\mathbf{n},t)= P(n_1,n_2,\dots,t)= P(\hat{\mathbf{n}},n_i,t)$,
where $i=1,2, \dots$ and $\hat{\mathbf{n}}$ denotes the vector of all $n$'s not explicitly written, see \textit{e.g.} Ref. \cite{Gardiner}.
For the continuous-time Markov processes of on-site births and deaths and unbiased random walk between neighboring sites, the master equation for $P(\mathbf{n},t)$ has the following form:
\begin{widetext}
\begin{eqnarray}
   \partial_t P(\mathbf{n},t) &=&  \sum_{i}\Big\{\lambda(n_i-1) P(\hat{\mathbf{n}},n_i-1,t)+\mu(n_i+1) P(\hat{\mathbf{n}},n_i+1,t) -[\lambda(n_i)+\mu(n_i)] P(\mathbf{n},t)\Big\}\nonumber \\
 &+&D_0\sum_{i} \Big\{(n_{i-1}+1) P(\hat{\mathbf{n}},n_{i-1}+1,n_i-1,t)+ (n_{i+1}+1) P(\hat{\mathbf{n}},n_i-1,n_{i+1}+1,t) -2 n_i P(\mathbf{n},t)\Big\}\,,\label{master1}
\end{eqnarray}
\end{widetext}
where we have returned to dimensional units.

\section{Small fluctuations}
Small fluctuations around the deterministic TWS can be described
by the (approximate) Fokker-Planck equation obtained,
via a truncated expansion,
from Eq.~(\ref{master1}), see \textit{e.g.} Ref. \cite{MS}:
\begin{eqnarray}
\partial_t P(\mathbf{q},t)&=&-\nu\sum\limits_{i}\frac{\partial}{\partial q_{i}}\, \left\{f(q_i)\, P-\frac{1}{2K}\, \frac{\partial}{\partial q_{i}}\,\left[g(q_i)\, P\right]\right\}\nonumber \\
&+&D_0\sum\limits_{i} \left(\frac{\partial}{\partial q_{i}}-\frac{\partial}{\partial q_{i+1}}\right)\bigg[\left(q_{i}-q_{i+1}\right)\, P \bigg.\nonumber \\
&+&\left.\frac{q_{i}+q_{i+1}}{2K}\left(\frac{\partial}{\partial q_{i}}-\frac{\partial}{\partial q_{i+1}}\right)\, P\right]\,.
\label{ps070}
\end{eqnarray}
Here $g(q)=\bar{\lambda}(q)+\bar{\mu}(q)$ is a rescaled \emph{on-site} diffusion coefficient in the space of population sizes. Instead of dealing
with the Fokker-Planck equation directly, we will consider the Langevin equation to which it
is equivalent. Reintroducing a continuous description in space, we obtain \cite{Ito}
\begin{eqnarray}
\partial_t q(x,t) &=& \nu f(q)+ D \partial_x^2 q+\sqrt{\frac{\nu g(q)\,h}{K}}\, \eta(x,t) \nonumber \\
&+&\partial_x \left[\sqrt{\frac{2 q(x,t)\, Dh}{K}}\, \chi(x,t)\right]\,,
\label{rw040}
\end{eqnarray}
where $\eta(x,t)$ and $\chi(x,t)$ are independent Gaussian noises which have zero means and are
delta-correlated both in $x$ and in $t$:
\begin{equation}
\left\langle \eta(x,t)\eta(x^{\prime},t^{\prime})\right\rangle=\left\langle \chi(x,t)\chi(x^{\prime},t^{\prime})\right\rangle=\delta(x-x^{\prime})\, \delta(t-t^{\prime}).
\label{rw060}
\end{equation}
The first two terms on the right hand side of Eq.~(\ref{rw040}) are the same terms as in the deterministic equation~(\ref{rateeq3}). The last two terms describe multiplicative noises coming
from the on-site birth and death processes (the third term), and from the random walk (the fourth term, which
has the form of a flux). Importantly, the magnitudes of the two noise terms are comparable to each other
in the front region. The flux term does not appear in the Langevin equations used in many papers for modeling fluctuations of the front position caused by \emph{external} noise \cite{Mikhailov2,Rocco}. In addition, for external noise the analogs of function $g(q)$ are chosen  \textit{ad hoc} \cite{Mikhailov2,Rocco}, whereas $g(q)$ in Eq.~(\ref{rw040}) is uniquely determined by the on-site birth and death rates.

Pechenik and Levine \cite{Levine} used the Doi-Peliti formalism \cite{Doi} to derive the Langevin equation for a microscopic model which includes processes $A\rightleftarrows 2A$ and random walk.  This model corresponds to propagation into \emph{unstable} state. However, the derivation would hold for
propagation into a metastable state as well. The Langevin equation of Ref. \cite{Levine} describes the evolution
of the system in terms of the dynamics of a sequence of Poissonian states \cite{Gardiner}. A possible advantage of their description
is that the random walk noise is ``absorbed" by the variables and does not appear explicitly. An advantage of Eq.~(\ref{rw040}) is that it provides
a direct description of the dynamics of the system in terms of the (rescaled) population size $q(x,t)$.

Rescaling the position coordinate $x$ and time $t$ by $l_D$ and $1/\nu$, respectively, we can rewrite Eq.~(\ref{rw040}) as
\begin{equation}\label{2}
  \partial_{t} q(x,t) = f(q) +  \partial_{x}^2 q(x,t) + N^{-1/2} R(x,t,q) \,,
\end{equation}
where
\begin{equation}\label{R}
   R(x,t,q)= \sqrt{g(q)}\, \eta(x,t)+\partial_{x}[\sqrt{2q}\, \chi(x,t)]\,,
\end{equation}
where $N=K l_D/h \gg 1$ is the typical number of particles
in the front region. The smallness of $N^{-1/2}$, the effective magnitude of the intrinsic noise, enables one to solve Eq.~(\ref{2}) perturbatively  around the deterministic TWS, see Refs. \cite{Mikhailov2,Rocco}.  One looks for a perturbative solution
\begin{equation}\label{5}
q(\xi_X,t)=q_0(\xi_X)+q_1(\xi_X,t)\,,\;\;\;|q_1|\ll q_0\,,
\end{equation}
where $\xi_X=x-c_0 t-X(t)$, and $X(t)$ describes the noise-driven displacement of the front position with respect to the deterministic TWS.  Going over to the variables $\xi_X$ and $t$ and linearizing Eq.~(\ref{2}) around $q_0$, we obtain
\begin{equation}\label{6}
   \partial_t q_1= \hat{{\cal L}} q_1 +\dot X q_0^{\prime}  + N^{-1/2} R(x,t,q_0) \,,
\end{equation}
where
\begin{equation}\label{L}
   \hat{\cal L}=\partial^2_{\xi}+c_0 \partial_{\xi}+f^{\prime}(q_0)\,,
\end{equation}
and we kept the old variable $x$ in the noise term.

$\dot{X}(t)$ can be found from the solvability condition of Eq.~(\ref{6}). To this end let us consider the eigenmodes of the operator  $\hat{\cal L}$:  $\hat{\cal L}\phi_n=a_n \phi_n$.
The zero mode $\phi_0(\xi)=q_0^{\prime}(\xi)$, for which $a_0=0$, is the Goldstone mode,  coming from the translational invariance of the deterministic TWS. Importantly, the lowest excited mode is gapped \cite{Mikhailov2,vanSaarloos}. We expand
$$
q_1(\xi_X, t)=\sum_n {b_n(t)\phi_n(\xi_X)} \,,
$$
and fix the definition of $X(t)$ by setting $b_0(t)=0$. Now we multiply Eq.~(\ref{6}) by $e^{c_0\xi} \phi_0(\xi)$
and integrate over $\xi$
from $-\infty$ to $\infty$. As the operator $e^{c_0\xi} \hat{\cal L}$ is Hermitian, the functions $\phi_n$
with different $n$ are orthogonal to each other with weight function $e^{c_0\xi}$, so upon normalization
\begin{equation}\label{7}
   \int_{-\infty}^{\infty} d\xi \phi_n(\xi)\phi_m(\xi)e^{c_0\xi}=\delta_{nm} \,,
\end{equation}
where $\delta_{nm}$ is the Kronecker's delta. As a result,
\begin{equation}\label{8}
   \dot{X} =-\frac{\int_{-\infty}^{\infty} d\xi \,q_0^{\prime}\, e^{c_0\xi} R(x,t,q_0)}{N^{1/2}\,\int_{-\infty}^{\infty} d\xi \,(q_0^{\prime})^2 \,e^{c_0\xi}} \,.
\end{equation}
The front diffusion coefficient $D_f$, in the frame moving with the velocity $c_0$, is given by the relation
\begin{equation}\label{9}
   D_f =\lim_{t\to \infty} \frac{\int_0^{t} dt_1 \int_0^{t} dt_2 \,\langle\dot X (t_1) \dot X (t_2)\rangle}{2 t}\,.
\end{equation}
Using Eqs.~(\ref{9}), (\ref{8}) and (\ref{R}) we obtain, after some algebra (see Appendix),
\begin{equation}\label{Dflin}
    D_f=\frac{D}{s_0N}\,,
\end{equation}
where
\begin{equation}
\label{s}
s_0= \frac{\left[\int_{-\infty}^{\infty} d\xi (q_0^{\prime})^2 e^{c_0 \xi}\right]^2}{\int_{-\infty}^{\infty} d\xi\,\Big\{ \frac{1}{2}\,(q_0^{\prime}e^{c_0\xi})^2\,  g(q_0)+q_0 \left[\left(q_0^{\prime} e^{c_0 \xi}\right)^{\prime}\right]^2\Big\}}
  \,.
\end{equation}
That is, in order to calculate the front diffusion coefficient one only needs two ingredients: the deterministic TWS $q_0(\xi)$ and the function $g(q)=\bar{\lambda}(q)+\bar{\mu}(q)$ evaluated on $q_0(\xi)$. As expected \cite{vanSaarloos03,Panja}, $D_f/D$ scales as $N^{-1}$, in stark contrast with the propagation into an \emph{unstable} state where $D_f/D \sim (\ln N)^{-3}$ \cite{Derrida06}. The first and second terms
in the denominator of Eq.~(\ref{s}) come from the birth-death noise and random walk noise, respectively. We note that the numerator and the first term
in the denominator appeared in the calculations of the
front diffusion coefficient for the case of \emph{external} fluctuations \cite{Mikhailov2,Rocco}.  In that case, however, $g(q)$ is unrelated to the on-site birth and death rates.
Importantly, for $f^{\prime}(q_*)<0$ and $f^{\prime}(0)<0$, all the integrals entering Eq.~(\ref{s}) converge.
\section{Wentzel–-Kramers–-Brillouin (WKB) approximation}
\emph{Large} fluctuations of the front position render the Fokker-Planck approximation (\ref{ps070}) and the ensuing Langevin equation (\ref{rw040})
inapplicable. However, one can still exploit the small parameter $1/N$ via
a WKB approximation, see \emph{e.g.}  \cite{MS,EK1,Dykman}, by making the ansatz $P(\mathbf{n},t)=\exp[-K S(\mathbf{q},t)]$
in the master equation (\ref{master1}). In the leading order in $1/K$, this brings about a Hamilton-Jacobi equation $\partial_t S+H(\mathbf{q},\partial_\mathbf{q} S)=0$. In the continuum limit in $x$, the Hamiltonian functional $H=\nu \int dx \,w$. Here
\begin{equation}
 w=H_0(q,p) - \partial_x q\, \partial_x p + q \left(\partial_x p\right)^2 \label{H21}
\end{equation}
is the rescaled Hamiltonian density, $x$ is rescaled by $l_D$, and
\begin{equation}\label{onsiteH}
    H_0(q,p)=\bar{\lambda}(q) \left(e^p-1\right) + \bar{\mu}(q) \left(e^{-p}-1\right)
\end{equation}
is the rescaled on-site Hamiltonian \cite{MS,EK1}. The last term in Eq.~(\ref{H21}) comes from the random walk noise. Note the identities $H_0(q,0)=0$, $\partial_p H_0(q,0)=f(q)$, $\partial_{qp} H_0(q,0)=f^{\prime}(q)$
and  $\partial_{pp} H_0(q,0)=g(q)$.

Instead of directly solving the Hamilton-Jacobi equation for $S(\mathbf{q},t)$, consider the Hamilton's equations of motion
\begin{eqnarray}
\partial_t q&=&\partial_p H_0(q,p) + \partial_x^2q -2\partial_x\left(q\partial_x p\right),
\label{p100}\\
\partial_t p&=&-\partial_q H_0(q,p) - \partial_x^2 p -\left(\partial_x p\right)^2
\label{p110}
\end{eqnarray}
for the population size $q(x,t)$ and effective momentum $p(x,t)=h\,\delta S/\delta q$.
Deterministic dynamics take place in the invariant hyperplane $p=0$, where Eq.~(\ref{p100}) reduces to Eq.~(\ref{rateeq3}). Fluctuations involve a non-zero $p(x,t)$, and this is
the case of our interest.
To have a well-defined problem for Eqs.~(\ref{p100}) and (\ref{p110}) one needs to specify the boundary conditions, both in space and in time. The boundary conditions in space correspond to zero-energy fixed points of the on-site Hamiltonian: $\{q(-\infty,t), p(-\infty,t)\}=\{q_*,0\}$ and $\{q(\infty,t), p(\infty,t)\}=\{0,0\}$.  The boundary conditions in time involve, in general, specifying kink-like population size profiles $q_1(x)$ and $q_2(x)$ at $t=0$  and $t=\tau$, respectively. After having solved the Hamilton's equations, one can calculate the action along the trajectory of the system in the functional phase space $q,p$:
\begin{equation}\label{action}
   K \mathcal{S}=N \int_{-\infty}^{\infty} dx \int_0^{\nu \tau} dt \;[p(x,t) \partial_t q -w]\,.
\end{equation}
We are interested in the probability of observing, in the frame moving with the velocity $c_0$, a
front displacement $X$ during a time interval which is much longer than the front relaxation time: $\nu \tau\gg 1$. Our crucial assumption is that this probability is described, in the leading order of $1/(\nu\tau) \ll 1$, by a traveling wave solution (TWS) of Eqs.~(\ref{p100}) and (\ref{p110}). This leads to two coupled ordinary differential equations for $q(\xi)$ and $p(\xi)$, where $\xi=x-ct$ and $c=X/\tau$ is the average front velocity during time $\tau$:
\begin{eqnarray}
  q^{\prime\prime}-2(q p^{\prime})^{\prime}+c q^{\prime}+ \partial_{p} H_0(q,p)&=& 0\,,\label{TWgenq}\\
  p^{\prime\prime}+(p^{\prime})^2-c p^{\prime}+ \partial_{q} H_0(q,p)&=& 0\,.\label{TWgenp}
\end{eqnarray}
The boundary conditions are $q(-\infty)=q_*$ and $q(\infty)=p(-\infty)=p(\infty)=0$.
As $\partial_t q = -c q^{\prime}(\xi)$ for a TWS, Eq.~(\ref{action}) reduces to
\begin{equation}\label{actionTWS}
   K \mathcal{S}(c)=-N \nu \tau \int_{-\infty}^{\infty} d \xi \;[cp(\xi)q^{\prime}(\xi)+w]\,.
\end{equation}

Remarkably, Eqs.~(\ref{TWgenq}) and (\ref{TWgenp}) possess a conservation law:
\begin{equation}\label{conservation}
H_0[q(\xi),p(\xi)]+q^{\prime} p^{\prime}-q(p^{\prime})^2=C=\mbox{const}
\end{equation}
where, by virtue of the boundary conditions, $C=0$.

Although especially suitable for dealing with large fluctuations of the front position, the WKB theory also captures typical, small fluctuations. To see it, let us linearize Eqs.~(\ref{TWgenq}) and (\ref{TWgenp}) around the deterministic TWS (\ref{4}):
\begin{equation}\label{linear}
q(\xi)=q_0(\xi)+u(\xi)\;\;\;\mbox{and}
\;\;\;c=c_0+\delta c\,,
\end{equation}
where $|u|\sim |p| \sim |\delta c| \ll 1$. We obtain
\begin{eqnarray}
 \hat{\cal L} u&=& -\delta c\, q_0^{\prime}-g(q_0) p + 2 (q_0 p^{\prime})^{\prime}\,,\label{TWlinq}\\
  \hat{\cal L}^* p&=& 0\label{TWlinp}
\end{eqnarray}
with zero boundary conditions at $\pm \infty$. The operator $\hat{\cal L}$ was defined in Eq.~(\ref{L}), whereas the operator
$\hat{\cal L}^*=d^2_{\xi}-c_0 d_{\xi}+f^{\prime}(q_0)$ is adjoint to $\hat{\cal L}$.

For the small deviations, the action (\ref{actionTWS}) can be expanded to second order.
The linear terms cancel out, whereas the quadratic terms yield, after some algebra,
\begin{equation}\label{actiongen}
   K \mathcal{S} (\delta c)= N \nu \tau \int_{-\infty}^{\infty} d\xi \left[\frac{1}{2} g(q_0) p^2+q_0(p^{\prime})^2\right]\,.
\end{equation}
As one can see, the knowledge of $p(\xi)$ alone is sufficient for evaluating the action (and the front
diffusion coefficient, see below) for small fluctuations. Therefore, let us consider Eq.~(\ref{TWlinp}).
One independent solution of this equation, $p_1(\xi)= e^{c_0 \xi} q_0^{\prime}(\xi)$, obeys the zero boundary conditions at $\pm \infty$ \cite{another}. Therefore, $p(\xi)=p_0 p_1(\xi)$ with an \textit{a priori} unknown constant $p_0$.
To determine this constant, we turn to Eq.~(\ref{TWlinq}) for $u(\xi)$, where $p(\xi)$ enters the effective forcing term
$$
F(\xi)= -\delta c\, q_0^{\prime}(\xi)-p_0 g[q_0(\xi)] p_1(\xi) + 2 p_0[q_0(\xi) p_1^{\prime}(\xi)]^{\prime}\,.
$$
Demanding zero boundary conditions at $\pm \infty$, one can obtain a solvability condition:
$\int_{-\infty}^{\infty} d\xi \, e^{c_0 \xi} q_0^{\prime}(\xi) F(\xi)=0$. This yields
\begin{equation}\label{p0gen}
 \frac{p_0}{\delta c}=-\,\frac{\int_{-\infty}^{\infty}d\xi (q_0^{\prime})^2\, e^{c_0 \xi}}{\int_{-\infty}^{\infty}  d\xi \big\{(q_0^{\prime} e^{c_0 \xi})^2  g(q_0) +2 q_0 \left[(e^{c_0\xi} q_0^{\prime})^{\prime}\right]^2 \big\}}\,.
\end{equation}
Then, from Eq.~(\ref{actiongen}), $ K \mathcal{S} = (1/4) s_0 N \nu \tau (\delta c)^2 $,
where $s_0$ is defined in Eq.~(\ref{s}), and $\delta c$ is still dimensionless.
Recalling that $\delta c = X/\tau-c_0$, we see that the probability distribution $P(X,\tau)$ is a Gaussian in the frame moving with the deterministic front velocity $c_0$. Normalizing this distribution to unity we obtain, already in the dimensional variables,
\begin{equation}\label{gauss}
    P(X,\tau)\simeq \left(\frac{s_0N}{4 \pi D \tau}\right)^{1/2} \exp \left(-\frac{s_0 N X^2}{4 D \tau}\right)\,,
\end{equation}
where $|X/\tau-c_0|\ll (\nu D)^{1/2}$. Equation~(\ref{gauss}) coincides with the Green function of a diffusion equation,  so the front diffusion coefficient $D_f=D/(s_0N)$, in a perfect agreement with Eq.~(\ref{Dflin}),
derived from the Langevin equation.

Rare large fluctuations of the front position correspond to $|p|\gtrsim 1$ in the WKB formalism. This demands a full nonlinear solution of Eqs.~(\ref{TWgenq}) and (\ref{TWgenp}) which in general is only accessible numerically.  Examples of such calculations are presented in the next section.

We should emphasize that our results on fluctuating \emph{extinction} fronts are conditional
on the absence of attempted extinction events in the region to the left of the front. This conditioning is especially important for $c_0<0$. Here an extinction front appears already in deterministic description, and formation of \emph{critical nuclei} of the state $q=0$ is possible \cite{Mikhailov,MS}.

\section{Example}
Consider three on-site reactions: $A \to 0$, $2A \to 3A $ and $3A \to2A$, with rate coefficients $\mu_0$, $\lambda_0$ and $\sigma_0$, respectively ~\cite{AM2010,MS,EK2,different}. It is convenient to perform the rescaling so that
$\bar{\lambda}(q)=2q^2$ and $\bar{\mu}(q)=\gamma\, q+q^3$, where $\nu=3\lambda_0^2/(8\sigma_0)$, $K=3\lambda_0/(2\sigma_0)$, and $\gamma=8\mu_0\sigma_0/(3\lambda_0^2)$.
The on-site dynamics exhibits bistability at $\delta^2 \equiv 1-\gamma>0$. In this case the zeros of function $f(q)=-q\,(q-q_u) (q-q_*)$
describe two stable fixed points of deterministic theory, $0$ and $q_*=1+\delta$, and an unstable fixed point $q_u=1-\delta$ such that $0<q_u<q_*$. The deterministic TWS $q_0(x-c_0 t)$ is \cite{Murray,Mikhailov,Petrovskii}
\begin{equation}\label{Q0}
    q_0(\xi)=\frac{\delta +1}{1+e^{(1+\delta)\,\xi/\sqrt{2}}}\,,\;\;\;\;
    c_0= \frac{3 \delta-1}{\sqrt{2}}\,,
\end{equation}
so the deterministic front is advancing at $1/3<\delta<1$, retreating at $0<\delta<1/3$ and stationary at $\delta=1/3$. The function $g(q)=\bar{\lambda}(q)+\bar{\mu}(q)=\gamma\, q+2q^2+q^3$.

\subsection{Small fluctuations}

\begin{figure}[ht]
\includegraphics[width=2.5 in,clip=]{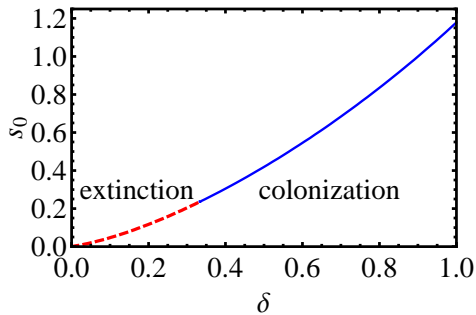}
\caption{(Color online) Function $s_0$, defined in Eq.~(\ref{s}), versus $\delta$ for
processes $A \to 0$, $2A \rightleftarrows 3A$, and random walk. The solid and dashed segments
distinguish between the colonization and extinction fronts, respectively.}
\label{s(delta)}
\end{figure}

Within theory of small fluctuations, one only needs to calculate $s_0=s_0(\delta)$ from Eq.~(\ref{s}) which enters the front diffusion coefficient $D_f=D/(s_0N)$.
All the integrals can be evaluated analytically. The result is a bit cumbersome, however, so we only present it graphically, see Fig. \ref{s(delta)}, and give its asymptotes:
\begin{equation}\label{asymptotes}
s_0(\delta)=\left\{\begin{array}{ll}
\frac{2 \sqrt{2}\,\delta}{9}+\frac{178 \sqrt{2}\, \delta^2}{135}\,, & \mbox{$0<\delta\ll 1$}\,, \\
\\
\frac{5\sqrt{2}}{6}-\frac{95 \sqrt{2} \,(1-\delta)}{72}\,, & \mbox{$0<1-\delta\ll 1$\,.}
\end{array}
\right.
\end{equation}
The latter asymptote corresponds to $\mu_0\sigma_0\ll \lambda_0^2$. For $\delta = 1/3$ (when the deterministic front is stationary) we obtain $s_0=\sqrt{2}/6$.

As one can see from Fig.~\ref{s(delta)}, the front diffusion $D_f=D/(sN)$ for the extinction fronts is stronger than for the colonization fronts. Furthermore, $D_f$ diverges as $1/\delta$ when $\delta \to 0$. This happens because, in this case, the integral in the denominator of Eq.~(\ref{s}) diverges at $\xi=-\infty$. The divergence comes from the first term of the integrand,
describing the birth-death noise. Notice that $\delta=0$ is the (saddle-node) bifurcation point of this model, and it is hardly surprising that fluctuations are enhanced in the vicinity of this bifurcation. Actually, very close to the bifurcation point
our perturbation theory breaks down. Indeed, it is only valid when the typical value of $\dot{X}$ is much less than the deterministic front velocity $\sim (\nu D)^{1/2}$. By virtue of Eqs.~(\ref{8}) and (\ref{R}), this boils down to a strong inequality $N \delta \gg 1$.

On the other hand, $D_f$ behaves regularly at $\delta=1$, that is when
the reaction $A\to 0$ is absent.  Here $s_0$ takes its maximum value $5\sqrt{2}/6$, so $D_f$ reaches
its minimum for this model. Notice that, for the reactions $2A\rightleftarrows 3A$ the state $q=0$ is marginally stable, but unstable with respect to any finite perturbation $q>0$.

\subsection{Large fluctuations}

Now consider large deviations of the front position. We solved nonlinear equations (\ref{TWgenq}) and (\ref{TWgenp}) for the TWS by the shooting method \cite{shooting}. For given $\delta$ and $c$ we first establish analytically, by linearization, the asymptotic behavior of the solutions at large negative $\xi$, where $|q-q_*|\ll 1$ and $|p|\ll 1$. This
leaves us with two arbitrary constants instead of four. By fixing one of them, we fix the location of the solution in $\xi$, and are therefore left with a single shooting parameter. The conservation law (\ref{conservation}) is used for accuracy control.

\begin{figure}[ht]
\includegraphics[width=2.3 in,clip=]{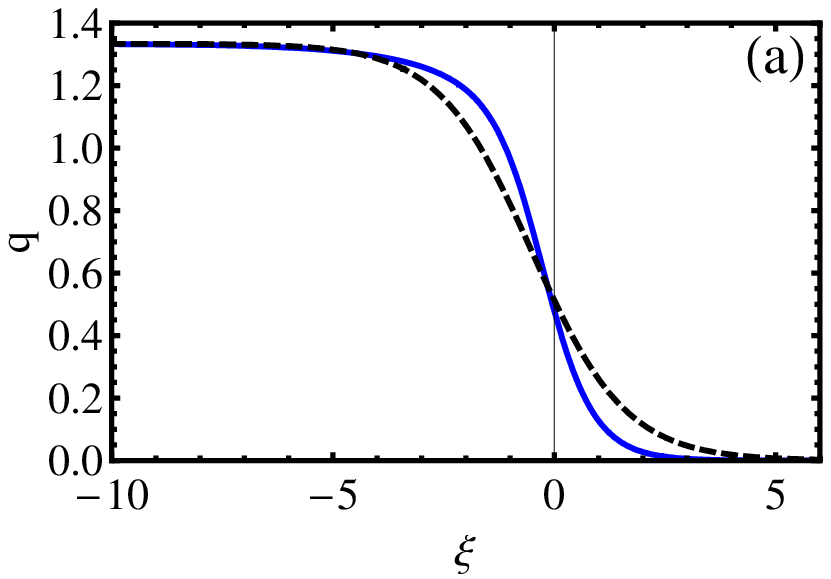}
\includegraphics[width=2.3 in,clip=]{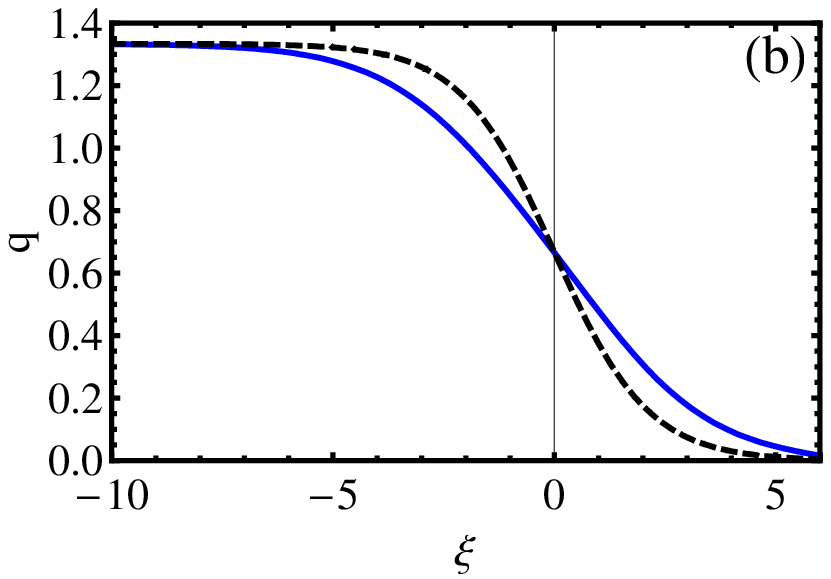}
\caption{(Color online) Solid lines: the most probable population size profiles for rescaled front velocities $c=2\sqrt{2}/3$ (a) and $c=-\sqrt{2}/3$ (b). Parameter $\delta=1/3$,  so that the deterministic front (dashed line) does not move.}
\label{q}
\end{figure}

Figure \ref{q} which shows the numerically found $q(\xi)$ profile for $\delta=1/3$ and $c=2\sqrt{2}/3$ (a) and $c=-\sqrt{2}/3$ (b). Recall that
the deterministic front is stationary in this case, so such high front velocities demand large fluctuations. The $q(\xi)$ profile -- the most probable population size profile for this front velocity -- is steeper than
the deterministic profile in Fig. \ref{q}a, and less steep that the deterministic profile in Fig.  \ref{q}b.

\begin{figure}[ht]
\includegraphics[width=2.5 in,clip=]{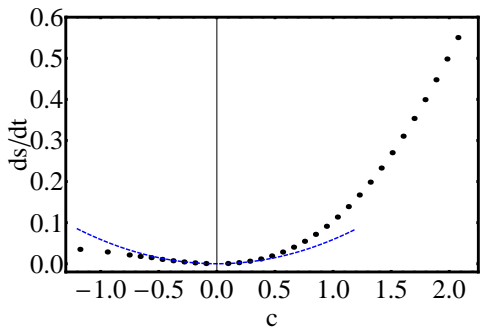}
\caption{(Color online) Symbols: numerically found ``action accumulation rate" $ds/dt$, see Eq.~(\ref{accumrate}), as a function of the rescaled front velocity for $\delta=1/3$. Dashed line: the prediction of linear theory, $ds/dt= \sqrt{2}\, (\delta c)^2/24$.}
\label{numaction}
\end{figure}

Figure \ref{numaction} shows the numerically found rescaled ``action accumulation rate"
\begin{equation}\label{accumrate}
   \frac{ds}{dt}=-\int d\xi (c p q^{\prime}+w)
\end{equation}
[see Eq.~(\ref{actionTWS})] versus $c$ for $\delta=1/3$. For small $|c|$ the numerical results agree with the linear theory prediction $ds/dt=\sqrt{2}\, (\delta c)^2/24$.  The tails of the distribution $\sim \exp[-K\mathcal{S}(c)]=\exp(-N \nu \tau ds/dt)$, however, are strongly non-Gaussian: under-populated for fluctuations pushing the front to the right (toward colonization), and over-populated for fluctuations pushing the front to the left (toward population extinction). That is, there is a much higher probability, at $c_0=0$, to observe an unusually fast extinction front, than an unusually fast colonization front.  This feature, however, can be model-dependent.

\section{Implications for Population Extinction}

Now let the spatial system size be large but finite: $L\gg l_D$, with impenetrable (reflecting) boundaries
at $x=0$ and $x=L$.  We still assume bistability, with  $q=q_*$ and $q=0$ being two linearly stable fixed points of the deterministic description. The zero population size throughout the system is an \emph{absorbing state} of the stochastic dynamics and, in the absence of immigration, the population ultimately reaches this state with probability one.

Let at $t=0$ a population front be located at $x=x_0$, where
$0<x_0<L$ and not too close to the boundaries. As before, we assume that the populated state $q=q_*$ is on the left. The \emph{mean time to extinction}, $\bar{T}$ can be very different depending on whether the deterministic front is retreating,  $c_0<0$, or advancing, $c_0>0$. For $c_0<0$, the population goes extinct essentially deterministically, and $\bar{T}\simeq x_0/|c_0|$ scales as the system size $L$. For $c_0>0$ the front first populates the whole system, around $q=q_*$,  with probability close to 1. This happens on a time scale $(L-x_0)/c_0$. Then the probability that the population goes extinct by time $t$ grows with $t$ very slowly, as $\bar{T}$ is expected to be \emph{exponentially} long  \cite{MS,EK1}.

Now, what happens in the special case of $c_0=0$ \cite{shift}? Here the front position wanders because of the intrinsic noise. As a result, there are two routes to extinction. In the first route the wandering front first reaches the left boundary $x=0$ where the population gets extinct rapidly. In the second route, the front first reaches $x=L$. Here the population develops a long-lived quasi-stationary distribution around $q=q_*$, and the probability to go extinct by time $t$  grows with time very slowly \cite{MS,EK1}.

The probabilities $p_l$ and $p_r$ of each of the two routes to occur, and the corresponding \emph{conditional} mean times to extinction $\bar{T}_l$ and $\bar{T}_r$ are described, in the leading order, by a classical first passage problem for random walk (of the front) in an interval \cite{Gardiner,Redner}. Therefore,  we immediately obtain $p_l=1-x_0/L$ and $p_r=x_0/L$. For the  extinction via the left boundary, $\bar{T}_l$ coincides, in the leading order, with the average first-passage time. The latter,
\begin{equation}\label{MTE}
    \bar{T_l}\simeq\frac{L x_0}{D_f}\left(2-\frac{x_0}{L}\right)\,,
\end{equation}
is well known, see \emph{e.g.} Ref. \cite{Redner}, p. 47. Typically, $\bar{T_l}\sim L^2/D_f$ in this case.  For the extinction via the right boundary, the conditional mean time to extinction $\bar{T}_r$ is expected to be exponentially long: much longer than the first-passage time for that boundary.

\section{Summary}

We have developed a systematic theory
of velocity fluctuations of a population front propagating into a metastable state. These fluctuations result
from the intrinsic noises of reactions (births, deaths, \textit{etc}.) and random walk, and we have identified the contributions of each of these noises to the velocity fluctuations. For typical, small fluctuations we have derived, from the underlying microscopic model, the Fokker-Planck equation and the effective Langevin equation. We have calculated the front diffusion coefficient using two different methods. The first method involved a perturbative solution of the Langevin equation in the weak-noise-limit, $1/N \ll 1$. The second method employed a perturbative solution of WKB equations in the form of a traveling front which involves both the rescaled population size $q(x,t)$ and the effective momentum $p(x,t)$. As expected, the front diffusion coefficient scales with $N$ as $1/N$. We have also evaluated, with a WKB theory, the probability of rare large fluctuations of the front. Here the probability distribution of the front position, in the frame moving with the deterministic front velocity, turns out to be
non-Gaussian and strongly asymmetric. We have also considered some implications of the theory for evaluating the population extinction risk. Our approach is readily extendable to multiple-step on-site processes such
as $n A\to (n+k)A$, where $n$ and $k$ are integers, and $k$ is not too large \cite{EK1,EK2}. Finally, most of our results (one notable exception being
the extinction aspects considered in Sec. VI) remain valid when both of the two linearly stable fixed points of $f(q)$ describe populated states, as in the
spruce budworm model, see \emph{e.g.} Ref. \cite{Murray}, p. 7.

\subsection*{Acknowledgments}
We thank David A. Kessler and Debabrata Panja for useful discussions. This work was supported by the Israel Science Foundation (Grant No.
408/08), by the U.S.-Israel Binational Science Foundation
(Grant No. 2008075), and by the Russian Foundation for Basic
Research (Grant No. 10-01-00463).

\section*{Appendix}
\renewcommand{\theequation}{A\arabic{equation}}
\setcounter{equation}{0}

Here we give some detail on the algebra leading to Eq.~(\ref{Dflin}). Substituting Eq.~(\ref{8}) in Eq.~(\ref{9}), we obtain
\begin{widetext}
\begin{equation}\label{A1}
D_f  = \lim_{t\to \infty} \frac{\int_{-\infty}^{\infty} d\xi_1 \int_{-\infty}^{\infty} d\xi_2 \int_0^t dt_1 \int_0^t dt_2 \,q_0^{\prime}(\xi_1)  q_0^{\prime}(\xi_2) e^{c_0(\xi_1+\xi_2)} \langle R(x_1,t_1,q_0)R(x_2,t_2,q_0)\rangle}{2 N t\left[\int_{-\infty}^{\infty} d\xi (q_0^{\prime})^2 e^{c_0\xi}\right]^2}\,.
\end{equation}
\end{widetext}
Now we perform the ensemble averaging using Eq.~(\ref{R}). As $\eta$ and $\chi$ are uncorrelated, we are left with two terms: $D_f=D_{f1}+D_{f2}$. The first term comes from the intrinsic noise of births and deaths:
\begin{widetext}
\begin{equation}\label{A2}
D_{f1} =\lim_{t\to \infty} \frac{\int_{-\infty}^{\infty} d\xi_1 \int_{-\infty}^{\infty} d\xi_2 \int_0^t dt_1 \int_0^t dt_2 \,q_0^{\prime}(\xi_1)  q_0^{\prime}(\xi_2) e^{c_0(\xi_1+\xi_2)} \sqrt{g[q_0(\xi_1)] g[q_0(\xi_2)]} \delta(\xi_1-\xi_2)\delta(t_1-t_2)}{2 N t \left[\int_{-\infty}^{\infty} d\xi (q_0^{\prime})^2 e^{c_0\xi}\right]^2}
\end{equation}
\end{widetext}
which yields
\begin{equation}\label{A3}
D_{f1} = \frac{\int_{-\infty}^{\infty} d\xi (q_0^{\prime} e^{c_0 \xi})^2  g(q_0)}{2N \left[\int_{-\infty}^{\infty} d\xi (q_0^{\prime})^2 e^{c_0\xi}\right]^2}\,.
\end{equation}
Now we calculate the second term, coming from the random walk noise:
\begin{widetext}
\begin{equation}\label{A4}
D_{f2} = \lim_{t\to \infty} \frac{\int_{-\infty}^{\infty} d\xi_1 \int_{-\infty}^{\infty} d\xi_2 \int_0^t dt_1 \int_0^t dt_2 \,q_0^{\prime}(\xi_1)  q_0^{\prime}(\xi_2) e^{c_0(\xi_1+\xi_2)} \big\langle\partial_{\xi_1}\left[\sqrt {2 q_0(\xi_1)}\, {\chi}(\xi_1,t_1)\right]\partial_{\xi_2} \left[\sqrt {2 q_0(\xi_2)} \,{\chi}(\xi_2,t_2)\right]\big\rangle}{2 N t\left[\int_{-\infty}^{\infty} d\xi (q_0^{\prime})^2 e^{c_0\xi}\right]^2} \,.
\end{equation}
\end{widetext}
We have
\begin{eqnarray}
  &&\big\langle\partial_{\xi_1}\left[\sqrt {2 q_0(\xi_1)}\, {\chi}(\xi_1,t_1)\right]\partial_{\xi_2} \left[\sqrt {2 q_0(\xi_2)} \,{\chi}(\xi_2,t_2)\right]\big\rangle \nonumber\\
  &&= 2 \partial^2_{\xi_1,\xi_2}\left[\sqrt {q_0(\xi_1)\,q_0(\xi_2)}\,\langle {\chi}(\xi_1,t_1){\chi}(\xi_2,t_2)\rangle\right] \nonumber\\
  &&=2 \partial^2_{\xi_1,\xi_2}\left[\sqrt {q_0(\xi_1)\,q_0(\xi_2)}\,\delta(\xi_1-\xi_2) \delta(t_1-t_2)\right]\nonumber \\
  &&=2 \delta(t_1-t_2)\,\partial^2_{\xi_1,\xi_2}\left[q_0(\xi_1)\,\delta(\xi_1-\xi_2) \right] \nonumber \\
&&=-2\delta(t_1-t_2)\,\partial_{\xi_1}\left[q_0(\xi_1)\,\delta^{\prime}(\xi_1-\xi_2) \right]\,.
\label{A5}
\end{eqnarray}
The double integration over time of $\delta(t_1-t_2)$ yields $t$ which cancels with $t$ in the denominator. Integration over
$\xi_2$ gives
\begin{equation}\label{A6}
    -2\int_{-\infty}^{\infty} d\xi_2 \,q_0^{\prime}(\xi_2) e^{c_0 \xi_2} \delta^{\prime}(\xi_1-\xi_2)=
    -2 \left[q_0^{\prime}(\xi_1) e^{c_0 \xi_1} \right]^{\prime} \,.
\end{equation}
Now we calculate the integral over $\xi_1$:
\begin{equation}\label{A7}
    -2\int_{-\infty}^{\infty} d\xi_1\, q_0^{\prime}(\xi_1)  e^{c_0 \xi_1}
  \big\{ q_0(\xi_1) \left[q_0^{\prime}(\xi_1) e^{c_0\xi_1}\right]^{\prime} \big\}^{\prime}\,.
\end{equation}
Integrating by parts, we can rewrite this integral as $2\int_{-\infty}^{\infty} d\xi\, q_0(\xi) \left[(q_0^{\prime} e^{c_0\xi})^{\prime}\right]^2$.
Therefore,
\begin{equation} \label{A8}
 D_{f2} = \frac{\int_{-\infty}^{\infty} d\xi\,q_0 \left[\left(q_0^{\prime} e^{c_0 \xi}\right)^{\prime}\right]^2}
  {N\left[\int_{-\infty}^{\infty} d\xi (q_0^{\prime})^2 e^{c_0 \xi}\right]^2}\,.
\end{equation}
Summing up the contributions (\ref{A3}) and (\ref{A8}), we obtain $D_f=D/(s_0N)$, where $s_0$
is defined in Eq.~(\ref{s}), and we
have restored dimensions by reintroducing $D$.


\begin{thebibliography}{99}
\bibitem{another} Another independent solution of Eq.~(\ref{TWlinp}), $p_2(\xi) = p_1(\xi) \,\int_0^{\xi} e^{c_0 \xi^{\prime}} p_1^{-2} (\xi^{\prime}) d \xi^{\prime}$ diverges at $\xi\to \pm \infty$ and should be discarded.
\bibitem{shift} More precisely, we need to demand here that the \emph{average} velocity $\bar{c}$ of the \emph{stochastic} front be equal to zero. For fronts propagating into metastable states the difference $|\bar{c}-c_0|$ is small, but the actual dependence of $|\bar{c}-c_0|$ on $N$ is unknown.  It would be natural to expect that $|\bar{c}-c_0| \sim  N^{-\alpha}$, where $\alpha\geq 1$ \cite{Panja,Kessler}. However, it has been argued recently \cite{Khain} that $|\bar{c}-c_0|$ can be \emph{exponentially} small in $N$.
\end{thebibliography}
\end{document}